# Robust Fault Detection and Classification in Power Systems via Physics-Informed and Data-Driven Learning

**Biswash Basnet[1], Varsha Sen[2]**

Lane Department of Computer Science and Electrical Engineering, West Virginia University, Morgantown, USA

**E-mail:** [1]bb00126@mix.wvu.edu, [2]vs00039@mix.wvu.edu

**Abstract**

Electrical faults in power transmission systems can severely hinder grid stability, equipment safety, and operational reliability. Traditional protection schemes, especially distance relays, depend on apparent impedance calculations that vary with fault, making them prone to misclassification. This can result in relay overreach, underreach, or complete maloperation, especially CT/PT saturation, evolving faults, or high-impedance scenarios. These limitations highlight the need for adaptive, data-driven alternatives. This paper presents an intelligent fault detection and classification framework based on supervised machine learning techniques that overcomes these challenges. The framework's robustness was validated under varying training sizes and Gaussian noise levels, demonstrating consistent accuracy and generalization across diverse learning conditions. Our approaches directly learn the complex nonlinear mapping between three-phase voltage/current patterns and the associated fault type, without assuming fixed impedance paths like traditional protection schemes. Utilizing line voltages and currents, we extract a rich set of derived features to represent the distinguishing characteristics of six fault categories. Multiple models, such as Artificial Neural Networks (ANN), Support Vector Machines (SVM), Random Forests, XGBoost, and Long Short-Term Memory (LSTM) networks, Physics-Informed Neural Networks (PINN), are developed and assessed on SMOTE-balanced datasets. These models classify faults without fixed thresholds or fault loop assumptions, thus improving sensitivity and robustness. The supervised machine learning approaches bridge the gap between traditional impedance-based protection and intelligent, scalable, data-driven grid analytics and can be integrated into a wide area monitoring and control system. Among the evaluated models, the PINN achieved the highest fault detection accuracy of 99.86%, while sustaining 99.79% of multiclass classification accuracy on the clean dataset. The PINN sustains high accuracy under 2–5% noise and 1–60% training data, maintains millisecond-level inference, and, by embedding power system equations, provides physical interpretability absent in purely data-driven baselines, enabling practical real-time protection.

**Keywords:** Fault Detection, Fault Classification, Supervised Learning, Transmission Line Protection, LSTM, Artificial Neural Networks (ANN), XGBoost

## 1. Introduction

The reliable operation of electrical power systems is crucial for maintaining modern infrastructure and industrial productivity. Line-to-ground (LG), line-to-line (LL), double-line-to-ground (LLG), three-phase (LLL), and three-phase-to-ground (LLLG) disturbances severely disrupt power delivery. These faults cause equipment damage and even lead to cascading blackouts. Early and accurate detection and classification of these faults are very crucial for ensuring grid stability, reducing outage time, and preventing catastrophic failures in electrical power systems. Traditional protection mechanisms such as impedance-based relays and overcurrent relays often fail to detect high-impedance faults, evolving grid topologies, or multiple simultaneous faults. As highlighted in [1], the apparent impedance measured by distance relays can vary based on fault type and location, making accurate classification essential for reliable protection. Recent advancements in artificial intelligence, particularly machine learning (ML), have opened new horizons in fault analysis. In this context, supervised machine learning-based approaches have been widely adopted for detecting and classifying faults. Anwar et al. [2] demonstrated the effectiveness of ensemble models for robust classification under noisy conditions. Porawagamage et al. [3] reviewed recent challenges in ML-based protection and proposed strategies for better data representation and real-time decision-making. The study by Chen et al. [4] emphasizes the significance of feature extraction for fault classification models using different methods. Moreover, advanced techniques combining ANNs with signal processing or optimization methods have shown high accuracy under various grid conditions [5], [6]. Building on this trend, data-driven methods of detection and classification of faults have attracted growing interest among research scholars, leveraging algorithms such as Support Vector Machines (SVM), Random Forests (RF), XGBoost, Long Short-Term Memory (LSTM), and optimized Artificial Neural Networks (ANN), becoming popular. Such methods have exhibited improved performance, and when combined with engineered features, along with data balancing methods such as SMOTE [7], [8]. Recent advances also underline the importance of resilience-oriented intelligent frameworks for strengthening modern grid protection [9]. Most recently, Physics-Informed Neural Networks (PINNs), which incorporate physics loss into the learning process, enhance interpretability along with power system fault analysis robustness [10]. The above developments indicate the growing possibility of intelligent paradigms of learning for exact, reliable, and real-time protection in smart grid operations as well as planning of future transmission systems [11]. However, existing approaches are often limited in scope; many focus on a narrow set of fault categories, assume abundant clean data, or rely purely on data-driven learning without physical interpretability. These constraints reduce robustness and real-time applicability, leaving a gap for frameworks that integrate physics-informed constraints with supervised models.

This paper proposes a combined, supervised learning framework for fault detection and classification in the power system. The architecture utilizes voltage and current signals to train separate models for binary detection and multi-class classification. Six algorithms, ANN, LSTM, SVM, Random Forest, XGBoost, and PINN, are evaluated using consistently engineered features and statistical measures. The dataset combines field-based realism with the variability of simulations to allow generalization. This framework allows for extensive benchmarking while maintaining modularity for real-time implementation. The goal is to enable scalable, automated, and accurate decision-making for wide-area monitoring, protection, and control (WAMPAC) across legacy power systems and smart grid deployments[12], [13].

## 2. Overview: Faults and Fault Types

### 2.1 Fault Types

Electrical power transmission systems are inherently vulnerable to a variety of disturbances, ranging from lightning strikes and insulation breakdowns to equipment aging and human error. These disturbances often cause accidental electrical connections or faults. Fault areas are potentially damaging phenomena requiring immediate attention for the system's stability, reliability, and safety. While traditional protection systems focus on detecting and isolating faults, modern networks demand more intelligent systems that can detect, classify, and localize faults swiftly to support dynamic relay coordination, situational awareness, and self-healing grid operations [14]. The most frequent is the single line-to-ground (LG) fault. It accounts for approximately (70–80)% of all transmission line faults. LG faults lead to high fault currents in the affected phase and a drop in its voltage, while the other phases may experience transient overvoltage, affecting insulation coordination and causing tripping misjudgments if not accurately classified [15]. Line-to-line (LL) faults, responsible for about (10–15)% of cases, are marked by high current exchange between the two involved phases and abnormal voltage conditions that can mimic more severe fault types if only magnitude-based detection is used. Double line-to-ground (LLG) faults make up the remaining (10–15)% and involve two phases shorted to the ground, producing large, unbalanced currents and the presence of zero-sequence components that significantly affect relay behavior. Though three-phase faults (LLL and LLLG) are less frequent, they can cause the most extensive system-wide damage due to symmetrical high fault currents and simultaneous voltage collapse in all phases [16].

### 2.2 Importance of Fault Identification and Classification

Understanding the effects of fault on system parameters is crucial for designing effective protection schemes [1]. For instance, LG introduces a low-impedance path to ground, causing a high fault current along with a simultaneous drop in voltage in the affected phase:

$$I_{fault} = \frac{V_{prefault}}{Z_{line}+Z_{ground}} \qquad (1)$$

where $Z_{ground}$ is often very low, resulting in a high fault current in that phase, while other phases remain relatively unaffected. In contrast, line-to-line (LL) faults create high current in the two involved phases, and are governed by the inter-phase impedance:

$$I_{LL} = \frac{V_{ab}}{Z_{ab}} \qquad (2)$$

Double line-to-ground (LLG) faults combine phase-to-phase and ground paths, leading to unbalanced currents and distorted system symmetry. These variations demand accurate fault classification for appropriate relay response. Misclassification or failure to identify the fault type can lead to incorrect relay operation, miscoordination, and delayed clearance, increasing the risk of cascading outages or widespread blackouts [17]. From a protection standpoint, the danger arises when relays interpret the wrong fault impedance, tripping prematurely (overreach) or failing to trip when necessary (underreach).

This challenge becomes more critical in distance relays, where the apparent impedance seen by the relay is used to estimate the distance to a fault. That impedance is calculated as:

$$Z_{app} = \frac{V}{I} \qquad (3)$$

However, the value of $Z_{app}$ varies not only with the fault location, but also with the type of fault. For example, a relay calibrated for a three-phase fault may significantly miscalculate the impedance when it encounters an LLG fault, leading to protection failure [1]. In pilot protection and differential protection schemes, accurate fault type knowledge enables better coordination between line terminals. This reduces unnecessary tripping and enhances system dependability. Fault classification also plays a vital role in self-adaptive protection systems, where protection parameters are dynamically adjusted based on fault characteristics. For instance, protection zones may be scaled using:

$$Z_{zone} = Z_{line}(1 + k_{fault}) \qquad (4)$$

where $k_{fault}$ is a small adjustment factor that depends upon characteristics like ground current. Fault classification facilitates real-time control, predictive maintenance, and better planning even in bidirectional DER-dense systems [2], [16]. In modern systems, detection initiates the alarm while classification ensures that the action is correct—smart classification is the key to future-inclusive, fault-tolerant protection.

## 3. Problem Formulation

Fault detection is critical to grid dependability, preventing cascading outages and enabling timely protective action. While traditional systems focused solely on detection, modern protection schemes require both detection and classification to support selective isolation and adaptive responses. The fault type, its transient behavior, and the urgency of protective action all depend on the nature of the fault and the phases involved. This paper formulates fault detection and classification as two related supervised learning tasks. The system under study is a simulated three-phase transmission line, where line voltages and currents are captured under both normal and faulted conditions. Each instance is transformed into a 13-dimensional feature vector using time-domain statistics such as average (Iavg, Vavg), range (Irange), standard deviation (Istd), magnitude (Imag), and zero-sequence components (I0, V0). All these features were selected with the aim of improving the discriminability of the model for complicated fault situations. Two data sets are used in the training and testing of the models. The first set is for the binary classification between fault and non-fault, while the second is for the multi-class classification with 4-bit ground and phase involvement representation (G, C, B, A). The realization is in line with practical requirements, wherein the nature of the fault affects impedance visible to the relays as well as zone estimation. Misclassification will lead to overreach, underreach, or miscoordination in distance protection [1].

The proposed architecture is targeting near-real-time implementation at high-resolution inputs using either PMUs or IEDs. Using engineered statistical features with interpretable models such as XGBoost, SVM, ANN, and PINN facilitates the implementation of high-performance, interpretable, and scalable decision-making that is sufficient for digital substation wide-area protection as well as control [18].

## 4. Methodology

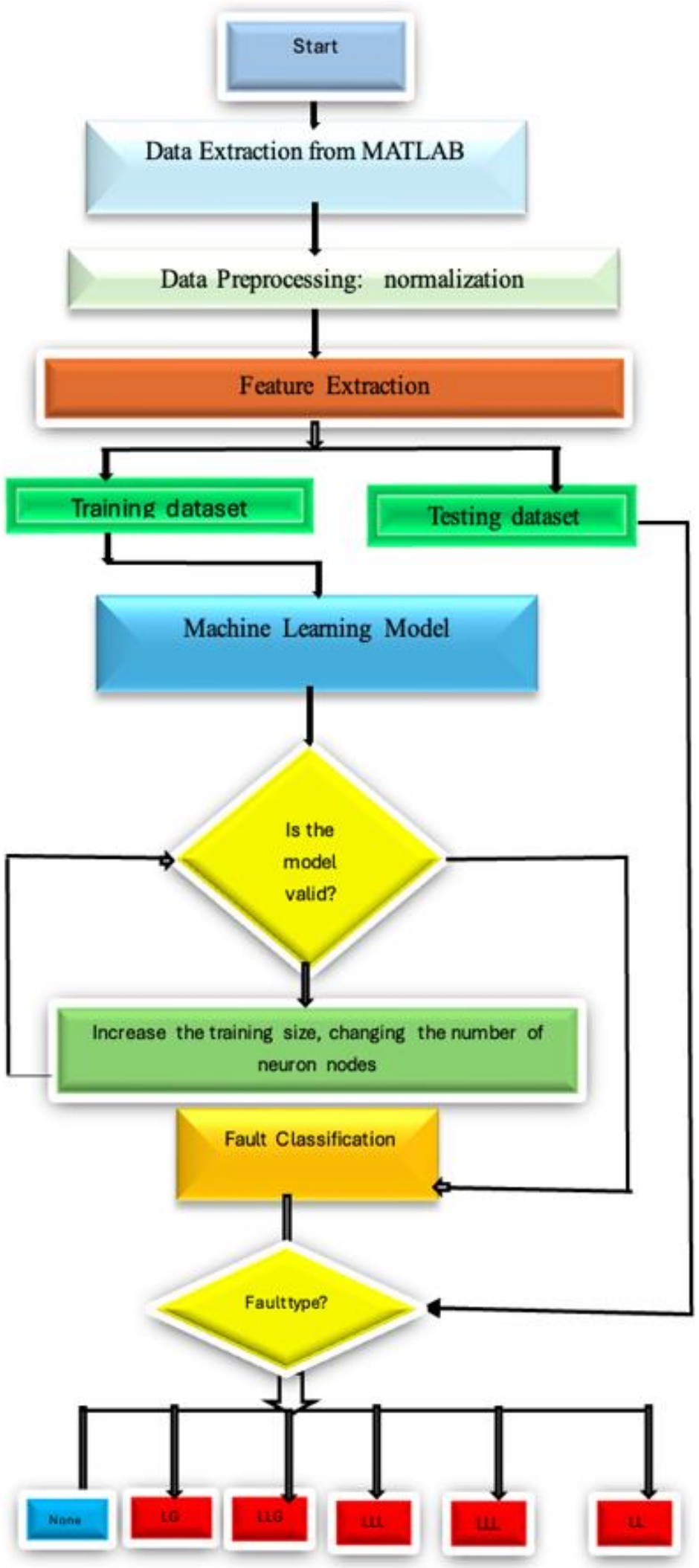

**Figure 1.** Flow Chart of Proposed Methodology

This section describes the proposed framework for intelligent fault detection and classification. It integrates data generation through simulation, domain-specific feature engineering, class balancing, and supervised learning using six state-of-the-art models. Figure 1 presents an overview of the proposed methodology, feature extraction, SMOTE balancing, and model training using six machine learning algorithms. It will be followed by performance evaluation under clean, noisy, and low-data conditions.

### 4.1 Data Generation

The IEEE 13-node test feeder has been employed to simulate realistic unbalanced distribution network behavior. Fault scenarios were simulated, including six conditions: no fault, LG, LL, LLG, LLL, and LLLG. Voltage and current signals (Va, Vb, Vc, Ia, Ib, Ic) were sampled at 10 kHz. Each instance was labeled using a 4-bit vector [G, C, B, A], where each bit represents the involvement of ground and phase conductors.

The datasets were modified by adding Gaussian noise to simulate real-world measurement imperfections. Initially tested up to 20%, noise levels were capped at 5% for model stability.

Each sample represents a 4–5 cycle window at 50 Hz. Z-score normalization has been used, as defined by:

$$x' = \frac{x-\mu}{\sigma} \qquad (5)$$

Stratified sampling was applied to construct training and testing sets with balanced class representation. The final dataset comprised approximately 10000 samples per class across the six fault categories, with variability introduced. The proposed framework is validated on the IEEE 13-node feeder, and can be extended to larger benchmark systems (e.g., IEEE 33-, 118-node) and real PMU datasets, as the feature extraction and learning modules scale independently of system size.

### 4.2 Feature Engineering and Balancing

Based on established practices in power system protection, and considering accuracy, a 13-dimensional feature vector was generated for each sample. These include Iavg, Vavg, Irange, Vrange, Istd, Vstd, Imag (RMS magnitude of currents), and approximated zero-sequence components $I_0$ and $V_0$. These features characterize fault signatures, including phase imbalance and asymmetrical behavior. For instance, high $I_0$ and $V_0$ values are typically associated with LG and LLG faults, while LLL faults remain symmetric [1].

For rare events like LLLG or LL faults, the Synthetic Minority Oversampling Technique (SMOTE) has been applied to address class imbalance. SMOTE generates synthetic samples for underrepresented classes and improves model recall, enhancing sensitivity without overfitting, while avoiding the instability of random oversampling and excessive noise of adaptive methods such as ADASYN.

**4.3 Machine Learning Models**

Six supervised machine learning models for fault classification and detection were constructed from the same standardized feature set, which had undergone SMOTE-balancing. The first one is the Artificial Neural Network (ANN), which was implemented with the three-layered Multi-layer Perceptron (MLP) network architecture. The binary detection network comprised two 256 and 128-neuron hidden layers, while the classification network comprised three deeper 256, 64, and 32-neuron layers. The stable convergence for both networks was achieved with ReLU activation functions along with early stopping techniques and adaptive learning rates. The ANN worked effectively by establishing the non-linear relationships between the feature dimensions.

Support Vector Machines with radial basis function (RBF) kernel were trained on these tasks. The hyperparameters, that is, the regularization coefficient (C) as well as kernel width($\gamma$), were optimized using grid search with three-fold cross-validation. The SVMs performed well in developing non-linear decision boundaries and consistently predicted fault types, with only slight variation due to feature changes.

Random Forest classifiers of 200 trees with 20 as the maximum tree depth were applied for classification in most scenarios. Their interpretability and generalizability were ensured with feature importance rankings that were inherent to them. RF generalized effectively to balanced and imbalanced categories of faults.

XGBoost classifiers were utilized for classification as well as for detection problems since they possessed the gradient boosting algorithm with regularization built into them. The model for detection possessed 100 estimators with a max depth of 5 and a 0.1 learning rate, while that of classification possessed 250 estimators with a depth of 8 and a 0.05 learning rate. XGBoost was found to possess steady resistance to noise as well as multicollinearity with extremely high accuracy for all the fault classes.

Long Short-Term Memory (LSTM) networks were used to learn temporal relationships from the waveforms of voltage and current. The input data streams were reshaped three-dimensionally to match LSTM input formats. The models included two stacked LSTMs with 64 units and 32 units, respectively, along with dense output with dropout regularization. LSTM models trained using the Adam optimizer and categorical cross-entropy error function were competitive in learning waveform dynamics and dependencies based on time.

Physics-Informed Neural Networks (PINNs) extended classical deep learning with the addition of physical constraints of Ohm's Law in the learning objective. Total loss is minimized in the learning Process.

$$\text{Total Loss} = \text{Data Loss} + \lambda \times \text{Physics Loss}$$

Here, the Data Loss corresponds to binary or categorical cross-entropy, while the Physics Loss represents the residuals of the three-phase Ohm's law equations. The weighting factor $\lambda$ is introduced to preserve the dominance of data-driven learning while enforcing physical consistency. In addition, dropout was applied from the outset to prevent overfitting, along with batch normalization to stabilize training. The block diagram for this is shown in Figure 2. The selected models reflect a balance between interpretability, computational feasibility, and prior adoption in fault analysis. SVM, RF, and XGBoost are established baselines in power system protection; ANN and LSTM capture nonlinear and temporal patterns; and PINN introduces physics-guided regularization to improve robustness. Other advanced architectures with very high parameter counts were not prioritized, as their training and deployment costs reduce suitability for real-time relays where fast and interpretable decisions are critical.

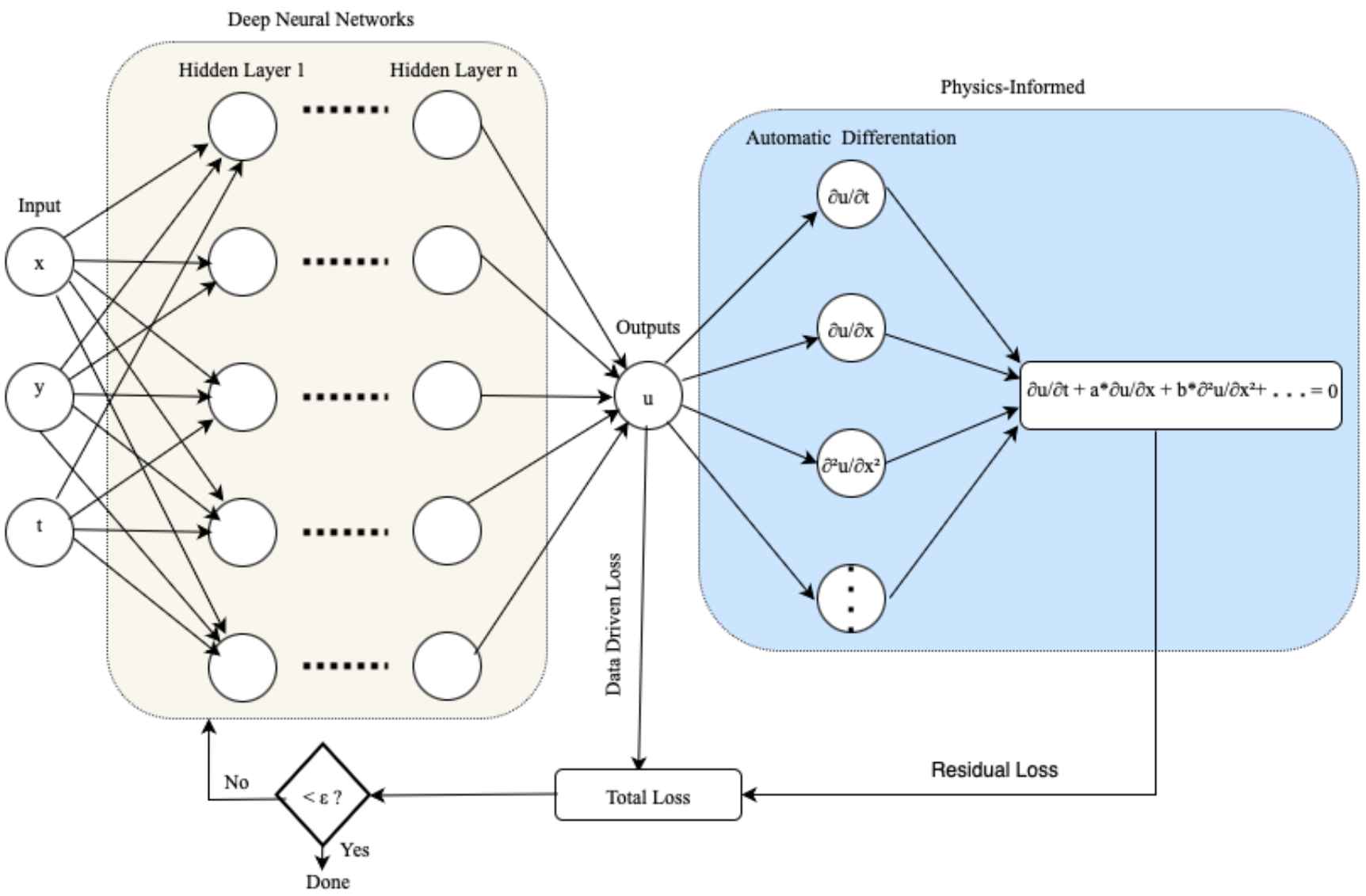

**Figure 2.** Physics-Informed Neural Network Block Diagram

### 4.4 Evaluation Metrics

Performance was assessed using accuracy, precision, recall, and F1-score, with macro-averaging to ensure a balanced judgment across all fault types. Mean Squared Error (MSE) and Mean Absolute Error (MAE) were also computed to capture confidence in probability outputs and stability in regression. Confusion matrices were analyzed to offer interpretation at the class level for model behavior. Models were tested on two challenging scenarios: variable train sizes (1—60%) and additive Gaussian noise (2–5%) that simulate data scarcity and uncertainty at the field level. These metrics were selected as accuracy captures overall correctness, precision/recall/F1 ensure reliability across minority fault types, and MSE/MAE quantify prediction stability under noisy or uncertain conditions, which are essential for protective relaying applications.

## 5. Results and Simulation

In this project, six machine learning models, ANN, LSTM, SVM, Random Forest (RF), XGBoost, and Physics-Informed Neural Network (PINN), are compared in this paper for fault detection and classification of transmission system faults. The performance of the models was compared based on three criteria: (i) accuracy on a clean dataset, (ii) generalizability with sparse data for training, and (iii) immunity to Gaussian noise (2–5%) during training. The performance measures utilized were accuracy, precision, recall, F1-score, Mean Squared Error (MSE), and Mean Absolute Error (MAE) in adherence to the real-time protective relaying requirements of smart grids.

### 5.1 Models' Performance under the Clean Dataset

All the models achieved higher-than 99% classification and detection accuracies with clean test data (Table 1). PINN achieved the highest detection accuracy of 99.86 % with the inherent physics-informed constraint enforcing generalizability. The optimal classification accuracy of 99.81% achieved in the case of SVM resulted from its ability to learn high-dimensional boundaries (Figure 3).

Table 1. Accuracy Comparison across Fault Detection and Classification Models

| Model | Fault Detection Accuracy | Fault Classification Accuracy |
|---|---|---|
| **ANN** | 99.34% | 99.74% |
| **LSTM** | 99.38% | 99.75% |
| **PINN** | 99.86% | 99.79% |
| **SVM** | 99.65% | 99.81% |
| **Random Forest** | 99.72% | 99.74% |
| **XGBoost** | 99.72% | 99.80% |

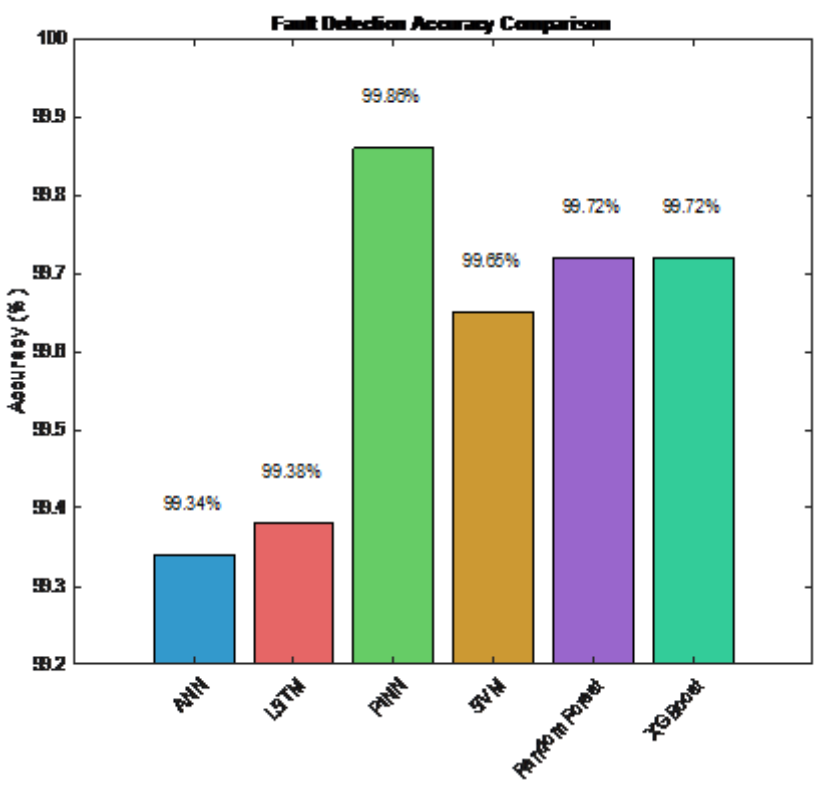

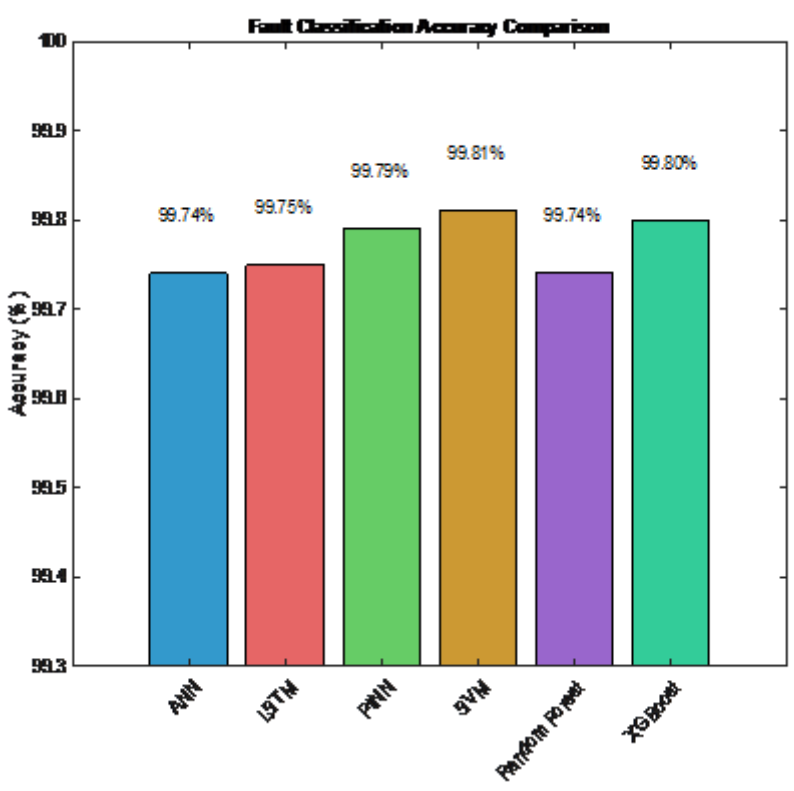

**Figure 3.** Bar Graphs of Accuracy across Fault Detection and Classification Models

## 5.2 MSE and MAE Error Analysis

Beyond accuracy, prediction stability was quantified using MSE and MAE (Table 2.). PINN recorded the lowest error values in both tasks (detection MSE: 0.0014, classification MSE:0.0210), as expected in high-confidence PINN prediction in an ideal setting (Figure 4). Low error rates in each class were observed in SVM and XGBoost, while ANN and LSTM recorded relatively higher variability.

Table 2. MSE and MAE across Fault Detection and Classification Models

| Model | Fault Detection MSE | Fault Classification MSE | Fault Detection MAE | Fault Classification MAE |
|---|---|---|---|---|
| **ANN** | 0.0066 | 0.0309 | 0.0066 | 0.0085 |
| **LSTM** | 0.0062 | 0.0293 | 0.0062 | 0.0081 |
| **PINN** | 0.0014 | 0.0210 | 0.0014 | 0.0061 |
| **SVM** | 0.0035 | 0.0197 | 0.0035 | 0.0057 |
| **RF** | 0.0028 | 0.0243 | 0.0028 | 0.0073 |
| **XGBoost** | 0.0028 | 0.0134 | 0.0028 | 0.0045 |

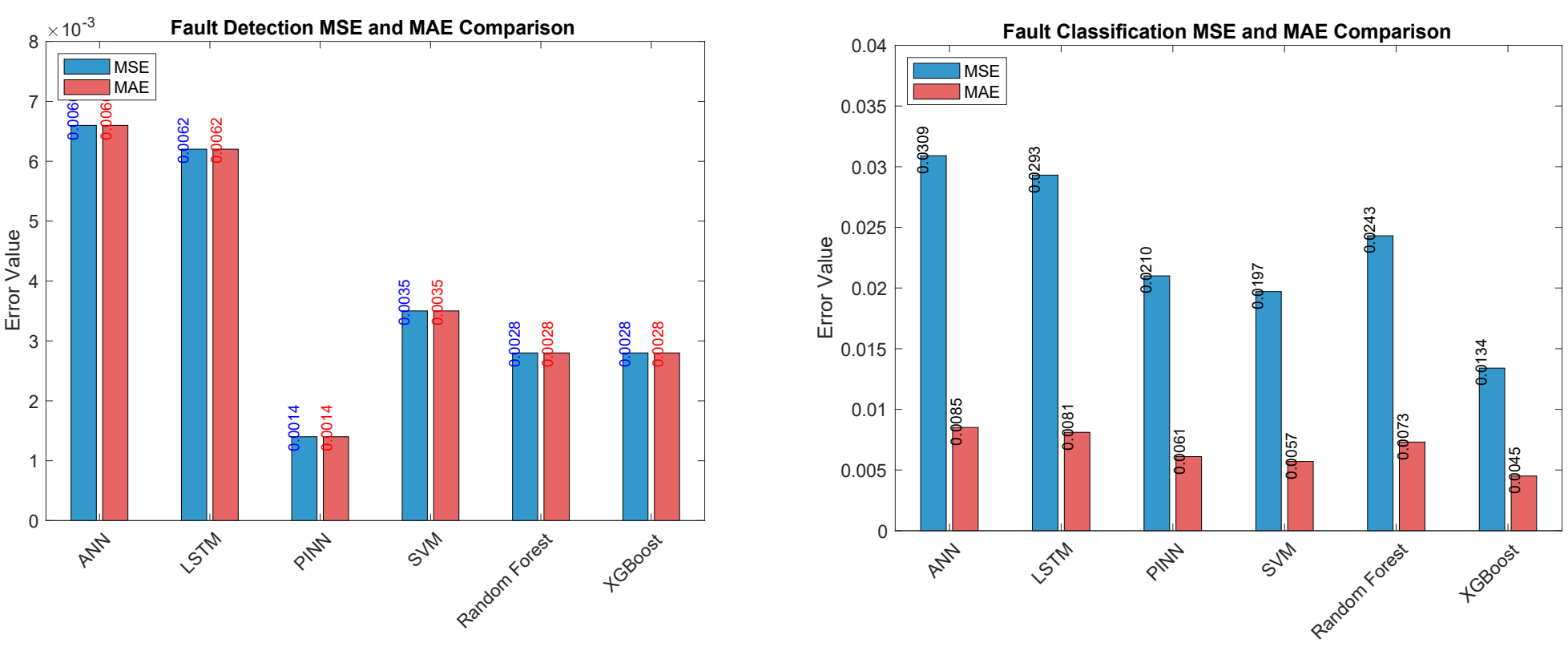

**Figure 4.** Bar Graph of MSE and MAE across Fault Detection and Classification Models

### 5.3 Models' Precision, Recall, and F1-Score

Macro-averaged F1-score, precision, and recall were calculated to evaluate class-wise reliability with particular interest in minority classes such as LLL and LLLG. All three metrics exceeded 0.99 for each model. PINN and SVM scored 1.00 regarding the three metrics that suggested highly balanced and exceedingly consistent classification among fault types (Table 3).

Table 3. Precision, Recall, and F1-Score Comparison Across Models

| Model | Precision | Recall | F1-Score |
|---|---|---|---|
| **ANN** | 0.99 | 0.99 | 0.99 |
| **LSTM** | 0.99 | 0.99 | 0.99 |
| **SVM** | 1.00 | 1.00 | 1.00 |
| **Random Forest** | 1.00 | 0.99 | 1.00 |
| **XGBoost** | 1.00 | 0.99 | 1.00 |
| **PINN** | 1.00 | 1.00 | 1.00 |

### 5.4 Confusion Matrix interpretation

The confusion matrices for the Physics-Informed Neural Network (PINN) demonstrate near-perfect generalization in both detection and classification tasks. The confusion matrices for PINN highlight that each matrix exhibits strong diagonal dominance with negligible off-diagonal entries, indicating minimal false positives and false negatives.

This level of accuracy reflects the model's ability to distinguish fault types reliably. The integration of physics-informed residual loss further reinforces this consistency by enforcing physical plausibility in predictions, thereby increasing confidence in the model's outputs for real-world protection applications. Compared to purely data-driven models, this physics-guided regularization reduces confusion between statistically similar fault signatures and improves generalization under noisy or limited training data, explaining PINN's consistent outperformance.

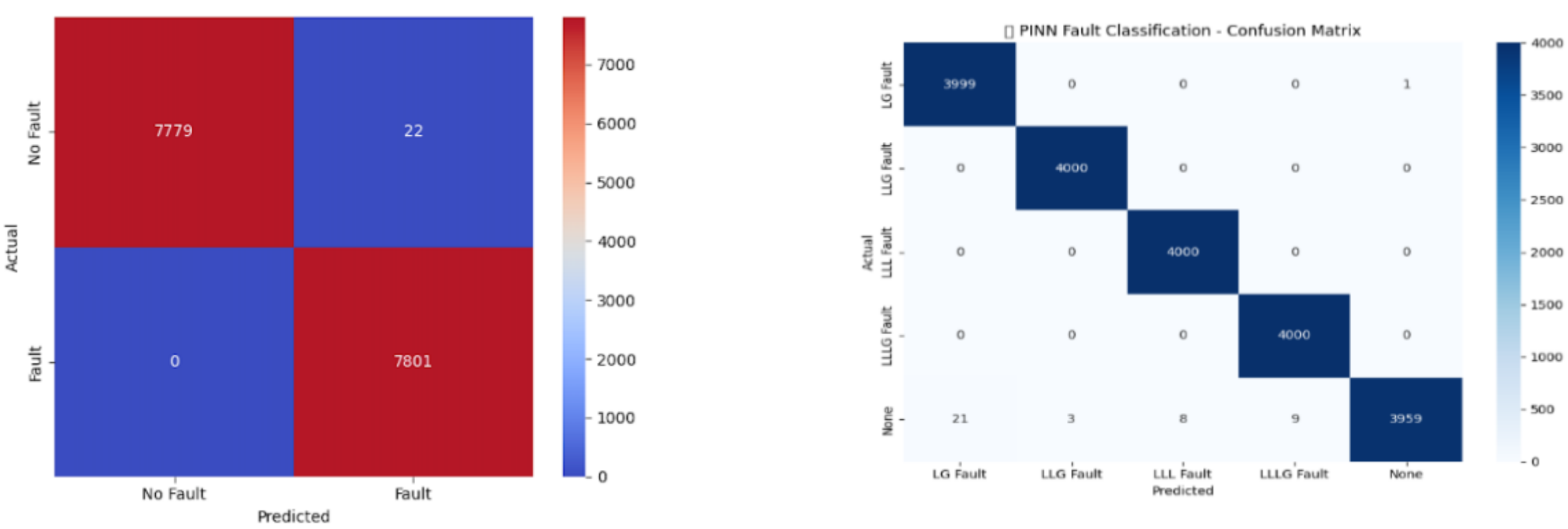

**Figure 5.** PINN Fault Detection and Classification Model Confusion Matrix

### 5.5 Models' Performance under Varying Training Size

Model performance across varying training sizes (1%—60%) revealed consistent trends. Detection and classification errors steadily decreased as training data increased, with most models reaching performance saturation around the 30% mark (Figure 6). PINN and XGBoost demonstrated strong generalization even with less than 20% training data, while SVM and Random Forest required larger datasets to stabilize. The slower stabilization of SVM and RF arises from their reliance on dense data distributions to establish decision boundaries, whereas XGBoost exploits boosting regularization and PINN leverages physics-informed constraints to achieve robust generalization with smaller training sizes. Accuracy curves for classification also exhibited saturation beyond 20% training data in most cases (Figure 7), highlighting the effectiveness of engineered features and learning capacity under constrained data availability.

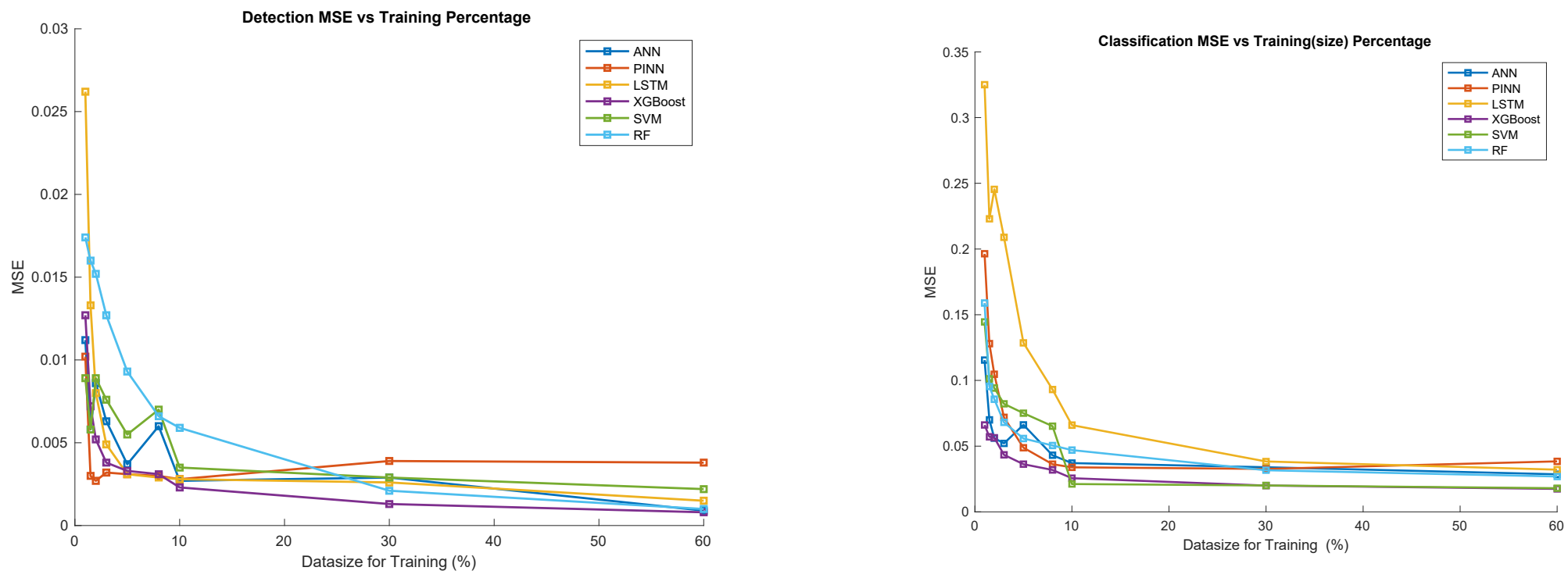

**Figure 6.** Fault Detection, Classification MSE vs Training Data size (%)

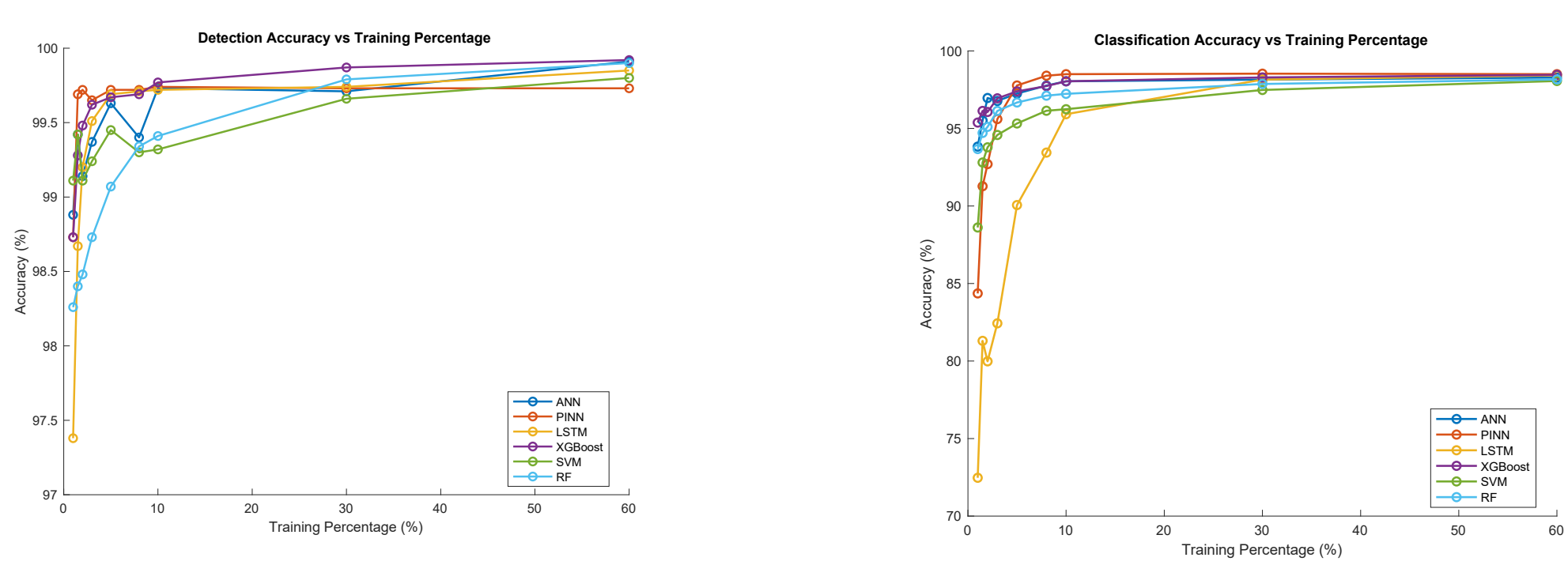

**Figure 7.** Fault Detection & Classification Accuracy vs Training Data size (%)

### 5.6 Model's Performance With Varying Noise Level

To assess robustness, all models were trained on data with 2-5% noisy data. Figures 8, 9, and 10 show accuracy comparisons across fault detection and classification tasks, highlighting each model's tolerance to noisy training conditions.

The Random Forest and SVM remained highly stable, showing minimal accuracy decay with increasing noise. ANN and PINN exhibited minor fluctuation but maintained overall reliable performance. Similarly, LSTM and XGBoost maintained stability in detection but experienced moderate fluctuation in classification accuracy.

In addition to accuracy under noise, computational feasibility was verified. ANN and LSTM incur higher training costs but yield millisecond-level inference. RF and XGBoost remain efficient with low memory overhead, while PINN adds physics-loss overhead during training, yet preserves inference latency comparable to ANN. These results confirm practicality for real-time PMU/IED-based deployment.

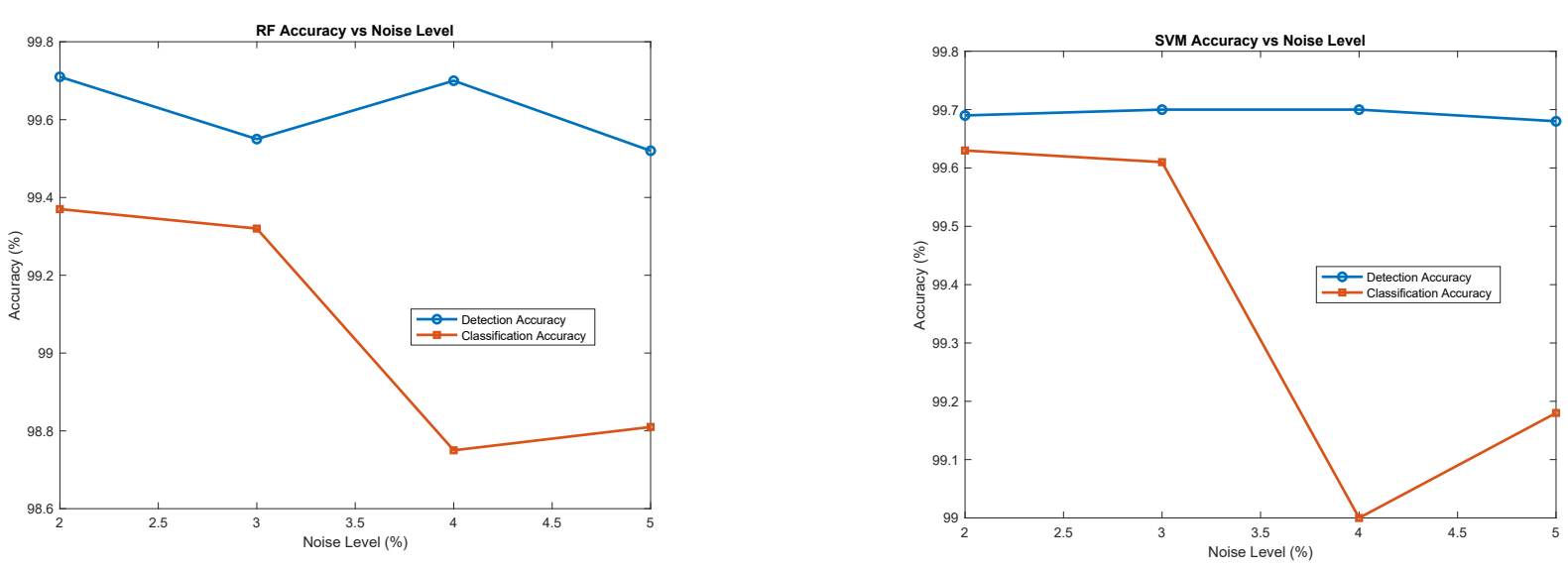

**Figure 8.** RF and SVM Accuracy Vs Noise Level

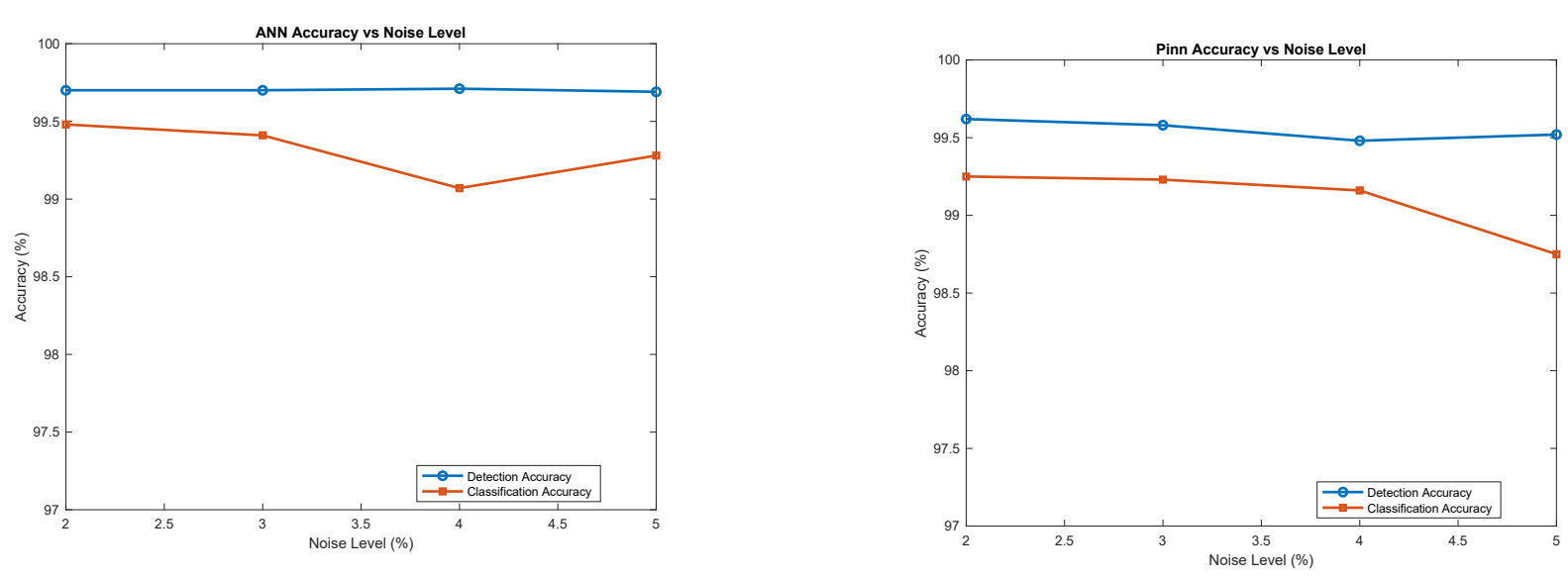

**Figure 9.** ANN and PINN Accuracy Vs Noise Level

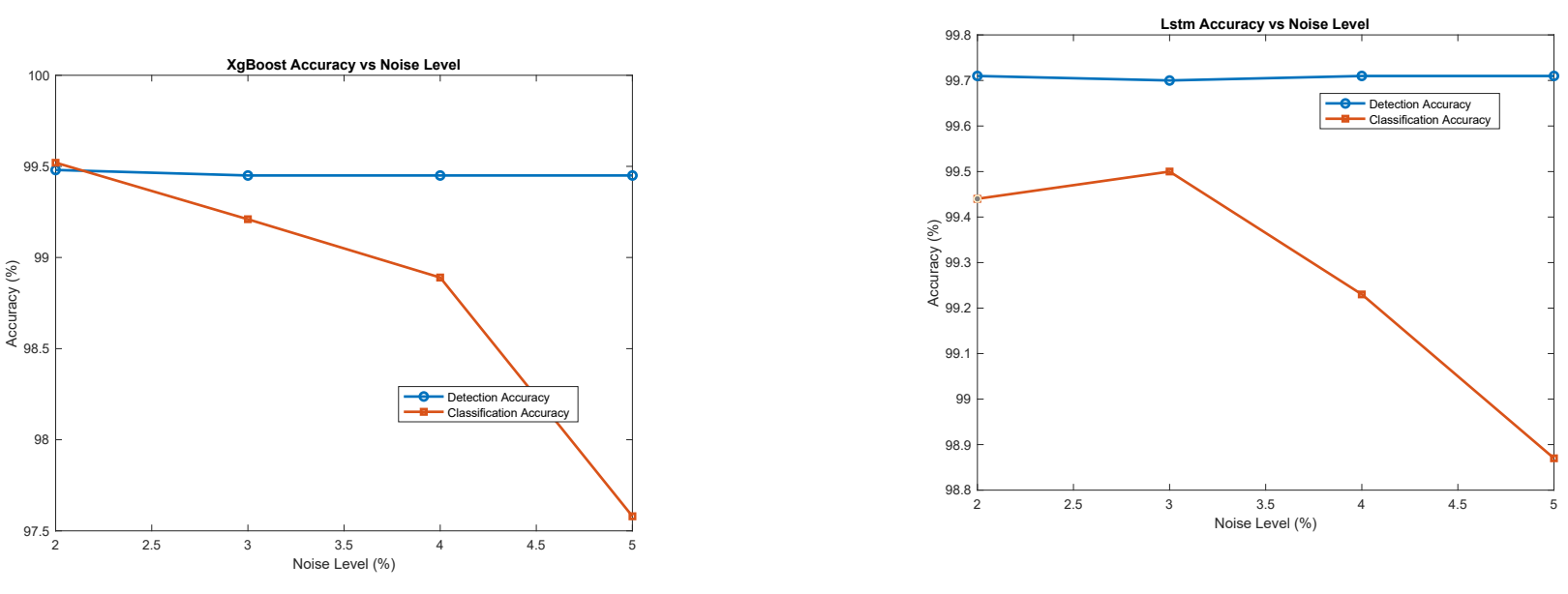

**Figure 10.** XGBoost and LSTM Accuracy Vs Noise Level

Across clean, reduced-data scenarios and different noise levels, the models consistently achieved high accuracy, precision/recall/F1 values exceeding 0.99, and error metrics confined within narrow ranges (Tables 1–3, Figs. 3–10). Taken together, all results and the uniformity of these results across multiple metrics and scenarios provide a strong indication of statistical validation and reproducibility of the reported findings, even though explicit confidence intervals are not included.

## 6. CONCLUSION AND FUTURE SCOPE

This work introduced a supervised learning framework for fault detection and classification in a power transmission system. Using time-domain voltage and current signals from IEEE 13-bus feeder simulations, we augmented with engineered features to train six models: ANN, LSTM, SVM, Random Forest, XGBoost, and a Physics-Informed Neural Network (PINN). All the models achieved above 99% accuracy with normal data, but PINN demonstrated particularly strong performance in terms of performance with limited data, as well as with noisy data. Since its loss function incorporated Ohm's Law, the PINN minimized confusion between identical types of faults and produced highly stable predictions. The comparison results also validated that ensemble methods, Random Forest, XGBoost, and sequence-aware models, LSTM, performed well, but that PINN performed with improved generalizability and interpretability, which is required for in-field use in protection relaying. The results show the advantage of embedding domain knowledge within learning frameworks to make the diagnosis of faults more reliable. In the future, we aim to extend the PINN architecture by introducing adaptive balancing of the losses, introducing convolutional or recurrent components, and dynamic physical models that extend beyond Ohm's Law. We will then evaluate the scheme on a realistic PMU dataset, leveraging transfer learning to mitigate latency and address missing values. We will then explore edge deployment techniques with compression, pruning, and explainable AI to enable rapid, interpretable performance in realistic smart grid cases.